\documentclass[prb,notitlepage,showpacs,twocolumn, superscriptaddress,aps,amsmath,amssymb,UFT8]{revtex4-1}

\usepackage[usenames,dvipsnames]{color} 
\usepackage{graphicx}
\usepackage{CJK}
\definecolor{LinkColor}{rgb}{0,0,.5}
\usepackage[colorlinks=true,linkcolor=BrickRed,citecolor=blue,urlcolor=LinkColor]{hyperref}

\newcommand{\ket}[1]{\left\vert{#1}\right\rangle}

\usepackage{pdfpages}

\makeatletter
\AtBeginDocument{\let\LS@rot\@undefined}
\makeatother

\begin{document}

\title{SARS-CoV-2 quantum sensor based on nitrogen-vacancy centers in diamond }

\author{Changhao Li}
\affiliation{ Research Laboratory of Electronics and Department of Nuclear Science and Engineering, Massachusetts Institute of Technology, Cambridge, MA 02139, USA}
\author{Rouhollah Soleyman} 
\affiliation{Department of Applied Mathematics, University of Waterloo, Waterloo, Ontario, Canada}

\author{Mohammad Kohandel}
\email[]{kohandel@uwaterloo.ca}
\affiliation{Department of Applied Mathematics, University of Waterloo, Waterloo, Ontario, Canada}
\author{Paola Cappellaro}
\email[]{pcappell@mit.edu}
\affiliation{ Research Laboratory of Electronics and Department of Nuclear Science and Engineering, Massachusetts Institute of Technology, Cambridge, MA 02139, USA}
\affiliation{Department of Physics, Massachusetts Institute of Technology, Cambridge, MA 02139, USA}


\begin{abstract}
 The development of highly sensitive and rapid biosensing tools targeted to the highly contagious virus SARS-CoV-2  is critical to tackling the COVID-19 pandemic. 
 Quantum sensors can play an important role, thanks to their superior sensitivity and fast improvements in recent years.
 Here we propose a molecular transducer designed for nitrogen-vacancy (NV) centers in nanodiamonds, translating the presence of SARS-CoV-2 RNA into an unambiguous magnetic noise signal that can be optically read out. We  evaluate the performance of the hybrid sensor, including its sensitivity and false negative rate, and compare it to widespread diagnostic methods
 . The proposed method is fast and promises to reach a sensitivity down to a few hundreds of RNA copies with false negative rate less than 1\%. 
 The proposed hybrid sensor can be further implemented with  different solid-state defects and substrates,  generalized to diagnose  other RNA viruses, and integrated with CRISPR technology.

\end{abstract}
\maketitle

\maketitle

\section*{Introduction} 
The rapid spread of the Coronavirus disease 2019 (COVID-19) has shown the importance for rapid and cost-effective testing of emerging new viruses. 
In the absence of specific drugs, early diagnosis and population surveillance of the virus load  are crucial to slow down and contain an outbreak. Unfortunately, as  of November 9, 2021, there has been more than 250 million reported COVID-19 cases with over 5 million deaths~\cite{WHO}. This huge cost in human life highlights the need for developing accurate testing for novel viruses with low false negative rate (FNR) and false positive rate (FPR), which enable taking appropriate preventive measures. Indeed, just the last 15 years have seen 4 pandemics (severe acute respiratory syndrome (SARS), Swine flu, Ebola, Middle East respiratory syndrome (MERS)) in addition to the severe acute respiratory syndrome coronavirus 2 -- SARS-CoV-2.

The diagnostic testing field for COVID-19 is rapidly evolving and improving in quality~\cite{KevadiyaNM2021}, but it has as yet been able to match the demand~\cite{KevadiyaNM2021,KucirkaAIM2020,WoloshinNEJM2020}. Diagnostics that is able to detect active infections are typically molecular-based, gauging the presence of the pathogen. The SARS-CoV-2 has a single-positive strand RNA genome, which remains in the body only while the virus is still replicating, and can be detected by various virology methods, most preeminently by the reverse transcriptase quantitative polymerase chain reaction (RT-PCR). Faster and more portable tests are provided by  rapid antigen tests  that  detect specific viral proteins found on the  virus surface, such as spike proteins. Unfortunately,  these faster tests are less reliable and cannot  quantify  the amount of virus. Serology tests, such as antibody tests, require a lag after infection and cannot catch the transmitting window.

While RT-PCR has been the primary method of  viral genome detection for SARS-CoV-2, it  has several drawbacks, as it requires trained personnel, special equipment, and careful design of the primer and probe. The samples need undergo RNA extraction as well as an amplification process which can take several hours, and might  degrade the diagnosis accuracy. The RT-PCR method also suffers from high FNR~\cite{KucirkaAIM2020,WoloshinNEJM2020}, which can be above 25~\% depending on the viral load of samples.  False negative results are especially consequential since they might lead to infected persons not  isolating and infecting others.  These challenges emphasize the imperative of finding a highly sensitive, accurate and rapid diagnosis method  for optimal diagnosis of COVID-19 and other viral outbreaks.

In recent years, quantum sensors~\cite{DegenRMP2017} have emerged as  powerful tools to detect chemical and biological signals. In particular,
nitrogen-vacancy (NV) centers in diamond act as stable fluorescence markers and magnetic field sensors, 
and have been investigated as quantum sensors for applications ranging from material science to chemistry and biology. A promising avenue to detect biological signals is to transduce them into magnetic noise -- using, e.g., magnetic molecules, such as gadolinium (Gd) complexes. Then, similar to FRET-based biosensors~\cite{RayJPCB2006}, the presence of a stimulus (in this case the virus) is detected as it induces a change in the NV fluorescence, following its separation from the magnetic nanoparticle. 
NV centers in nanodiamonds (NDs) have thus already demonstrated their ability to detect biological processes with high spatial resolution~\cite{MillerNature2020,LiNL2019,BarryPNAS2016,ChoiPNAS2020}, and are the object of intense  study due to their many favorable properties, ranging from very high photo- and thermal-stability to biocompatibility, in contrast to many florescent biomarkers in FRET-based approach.


Here, we introduce a quantum sensor based on NV centers in NDs that is capable to detect virus RNA, focusing on the SARS-CoV-2 virus. We show with numerical simulations that the proposed  sensor can detect as low as a few  hundreds of viral RNA copies in a one-second measurement time window when interacting with virus RNA sufficiently. Considering the distribution of relevant parameters  in realistic experiments,  we show that the sensor can reach a FNR less than 1 \% --considerably smaller than RT-PCR-- even without the RNA amplification process. The signal can be optically read out with a short acquisition time. The proposed diagnosis method has a low material cost and is scalable to simultaneous measurement of many samples.  We further discuss the potential integration of the proposed RNA quantum sensor with CRISPR technology to achieve even higher sensitivity, as shown in the supporting material~\cite{SOM}.  

While these ideas can be generalized to other RNA viruses (or more broadly any nucleic acid) and other solid-state defects, in the following we build a theoretical model of how the NV fluorescence is affected when the virus RNA cleaves the bound of Gd molecules with  the ND surface. We show how the quantum properties of the NV spin, which make it extremely sensitive to external perturbation, together to its nanoscale size combine to provide the exceptional performance of the proposed sensor.

\section*{Virus RNA sensor based on single NV centers}

\begin{figure*}
    \centering
    \includegraphics[width=0.9\textwidth]{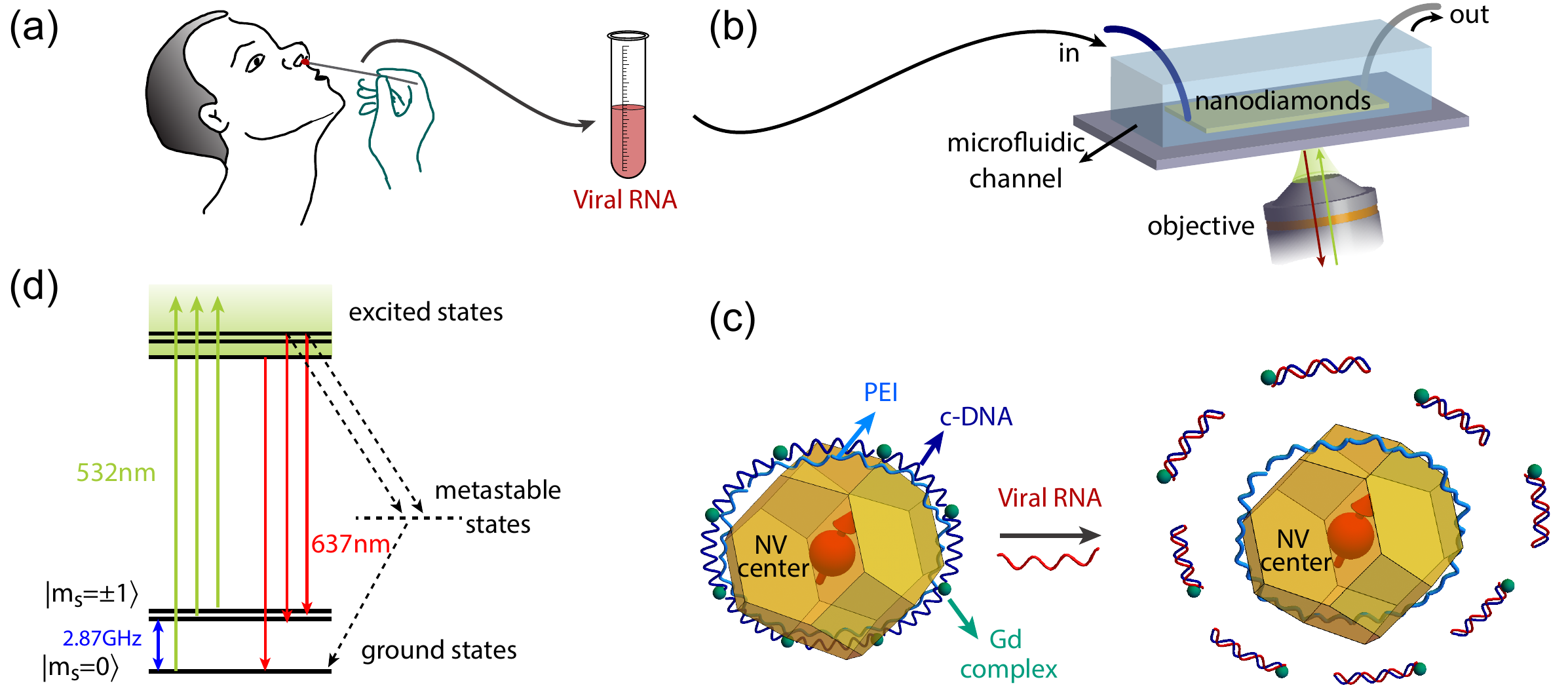}
    \caption{\textbf{Overview of the diagnosis protocol.}\textbf{(a)} A sample is collected from the upper respiratory tract, e.g., with a nasopharyngeal or throat swab~\cite{NasalSwabImage}, followed by nucleic acid extraction. \textbf{(b)}  Test samples that might contain virus RNA are loaded into microfluidic channels containing functionalized nanodiamonds. The emitted red fluorescence signal resulting from green laser excitation of the NV is collected via a confocal microscope or on a CCD. \textbf{(c)} Mechanism of magnetic noise quenching.   c-DNA is adsorbed on the surface of functionalized nanodiamond containing NV centers. The absorption is due to molecular interactions between  cationic polyethyleneimine (PEI) polymer on nanodiamond surface and c-DNA. Other polymers such as poly-l-lysine~\cite{Farrell2007} could also be used to bind the c-DNA sequences. A stable c-DNA fragment of SARS-CoV-2  can be  obtained by RT-PCR~\cite{XieCHM2020} or synthesis.  Gd$^{3+}$ complex molecules that can induce strong magnetic noise are connected to the c-DNA structure. In the presence of virus RNA, the base-pair matching of c-DNA and RNA leads to detachment of c-DNA-DOTA-Gd$^{3+}$ from the nanodiamond surface, resulting in weaker magnetic  interaction between Gd$^{3+}$ complex and NV centers inside the nanodiamond. \textbf{(d)} Energy level diagram of an NV center showing the optical transitions. Green laser  non-resonantly excites the NV spin to its excited state, and the NV decays back emitting red fluorescence photons. The $\ket{m_s=\pm1}$ states can also decay nonradiatively through the metastable singlet state and back to the $\ket{m_s=0}$ ground state  (dashed lines), providing a mechanism for both optical initialization to $\ket{m_s=0}$ and spin state-dependent optical readout.}
    \label{fig:Fig1_Scheme}
\end{figure*}

Our proposed diagnosis technique is based on the detection of viral RNA as  shown in Fig.~\ref{fig:Fig1_Scheme}(a). 
As in the standard RT-PCR method, viral RNA taken from an upper respiratory sample of patients is  isolated and purified using fast spin-columns. 
Contrary to RT-PCR,  reverse transcription is not be needed, as the sensing technique is based on direct detection of the virus RNA. Moreover, 
our sensor does not require  nucleic acid amplification  due to its high sensitivity to viral RNA, reducing the complexity, cost, and time of the test. The sample solution is directly ejected into microfluidic devices where the NV-based hybrid sensor is loaded beforehand. 

The negatively charged Nitrogen-Vacancy (NV) center is a point defect in the diamond lattice, consisting of a substitutional nitrogen atom with an adjacent vacancy. The NV electronic spin has a triplet ground state, with the spin levels  $\ket{m_s=0}$ and $\ket{m_s=\pm1}$ separated by a  zero-field splitting $\omega_0/2\pi = 2.87$ GHz. 
A short laser illumination can polarize the spin into the $\ket{m_s=0}$ level (Fig.~\ref{fig:Fig1_Scheme}), a state of effective low temperature, that can be recognized by its high fluorescence intensity. The system then relaxes to its thermal equilibrium state that displays a lower fluorescence intensity. The thermalization process occurs
in a characteristic time (the longitudinal relaxation time $T_1$)   dictated by the quantum spin magnetic environment, which can be indirectly affected by biological processes. 
In particular, paramagnetic molecules such as Gadolinium (Gd) complexes can induce strong transverse magnetic noise and significantly reduce the NV spin relaxation time~\cite{TetiennePRB2013,LiNL2019,Barton20}. 

As shown in Fig.~\ref{fig:Fig1_Scheme}(c), we non-covalently coat  NDs  by cationic polymers~\cite{PetrakovaNanoscale2016,Pandey16,Farrell2007,Krueger08c}, such as polyethyleneimine (PEI),  which can form  reversible  complexes  with  viral complementary DNA (c-DNA) sequences\cite{Boussif95} .
Magnetic molecules such as Gd$^{3+}$ complexes can be incorporated into the sequence~\cite{RotzACSnano2015}, forming hybrid c-DNA-DOTA-Gd pairs. For example, the amine modified end of the c-DNA is   able to bind to a Gd$^{3+}$ chelator such as 1,4,7,10-tetraazacyclododecane-1,4,7,10-tetraacetic acid (DOTA) through amide covalent bond~\cite{XuCC2011}.
 Due to the molecular electrostatic interactions between PEI and c-DNA-DOTA-Gd$^{3+}$, the Gd$^{3+}$ complexes will tend to  lie on the ND surface, in close proximity to NV centers, thus efficiently increasing the magnetic noise strength felt by NV spins and quenching their   $T_1$ time.

In the presence of viral RNA, the c-DNA-DOTA-Gd$^{3+}$ pair will detach from the ND surface due to  c-DNA and virus RNA hybridization. An analysis of the binding reactions~\cite{SOM} reveals that the formation rate of c-DNA-RNA hybrids is considerably higher than the binding between c-DNA and polymer-functionalized ND~\cite{YangAC2008,LiuCC2009}.

The newly-formed c-DNA-DOTA-Gd$^{3+}$/RNA compound will then diffuse freely in the solution, leading to an increased distance between Gd and ND. The NV centers  will feel weaker magnetic noise and have a longer $T_1$ time, indicated by a larger fluorescence intensity at fixed wait time. By optically monitoring 
the change in relaxation time, we can identify  the presence of virus RNA in the sample and even quantify the RNA  number. 
We note that, contrary to many other NV-based biological sensors~\cite{KucskoNature2013,SteinertNC2013,BarryPNAS2016,ChoiPNAS2020,MillerNature2020}, microwave driving is not necessary here in principle, thus reducing the complexity and cost of the experimental setup. To characterize NV charge state fluctuations that might mask the real $T_1$ decay,  fluorescence spectrum can be measured to extract the ratio of different charge states~\cite{Rendler2017,Grotz2012,Karaveli3938}.

\begin{figure*}
    \centering
    \includegraphics[width=0.95\textwidth]{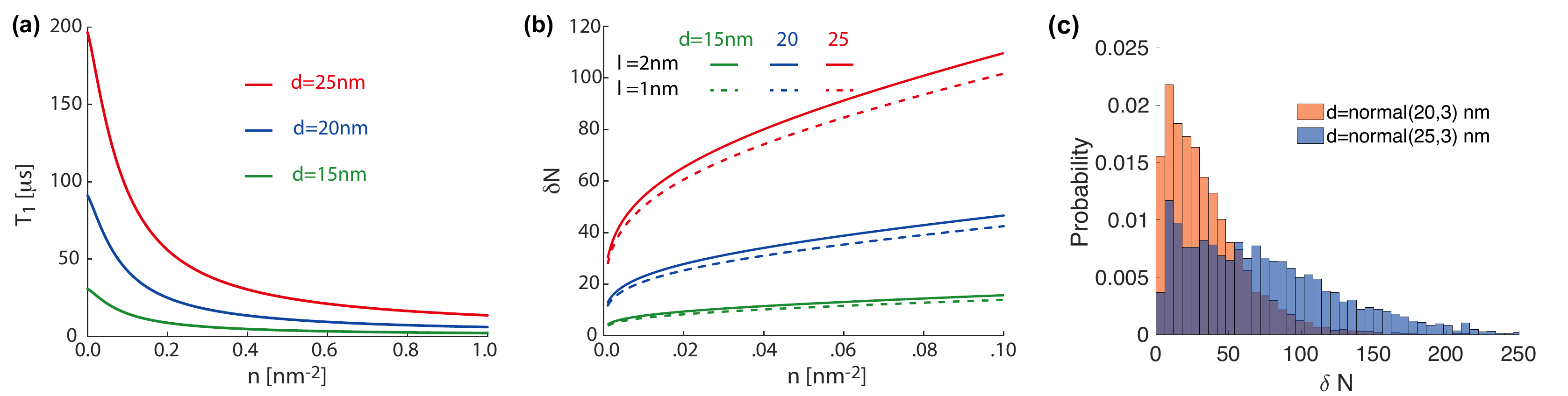}
    \caption{\textbf{Single NV sensor performance.} \textbf{(a)} Relaxation time $T_1$ as a function of the surface density of Gd$^{3+}$ complex molecules. The distance of Gd$^{3+}$ complexes and ND surface is $l=1$~nm. The plotted range of Gd density is  a very conservative estimate  based on previous studies of PEI ND coating~\cite{ZhangACSnano2009}, PEI-DNA binding~\cite{KimJPCL2012} and DNA-Gd attachment~\cite{RotzACSnano2015}. 
\textbf{(b)} Minimum detectable number of c-DNA-DOTA-Gd$^{3+}$ molecules in $T$=1~s integration time (i.e., quantum sensitivity) for a NV center at the ND origin.   
    Given the strength of the RNA-DNA hybridization compared to the DNA-PEI biding~\cite{SOM}, we assume freely diffusing Gd molecules after detachment. At that point, their  contribution to the $T_1$ can be safely neglected, as the distance between NV and Gd becomes significantly large~\cite{SOM}. 
\textbf{(c)} Probability density distribution of quantum sensitivity (for optimal  measurement time) considering the normal distribution of ND diameters, surface density of c-DNA-DOTA-Gd$^{3+}$ as well as position of NV spin in the ND.  The surface density $n$ follows a Normal (0.1,0.02)~nm$^{-2}$ distribution. The distance between Gd$^{3+}$ complex and ND surface is set to  $l$=1.5 nm and the NV center position is randomly sampled in the ND.  We fix the surface density of random paramagnetic centers at $\sigma=1$~nm$^{-2}$, so the contributions from the ND bulk spins and surface spins to the relaxation process remain constant.}
    \label{fig:Fig2_singleNV}
\end{figure*}

The proposed protocol relies on the magnetic noise induced by dipolar interactions between NV centers and Gd molecules. Here we introduce a simple model to describe the $T_1$ time  before and after the c-DNA/RNA hybridization and study the sensitivity of the proposed viral RNA sensor. 
We first consider for simplicity a single NV center at the center of a single ND with diameter $d$. 
The NV spin relaxation time is determined by the magnetic noise strength, with contributions from nuclear and electronic spins in the diamond bulk, paramagnetic defects on the ND surface, and Gd molecules that are connected by c-DNA:
\begin{equation}\label{eq:T1_formula}
    \frac{1}{T_1} \sim \frac{1}{T_{1,bulk}} + \gamma_e^2B_s^2 S_s(\omega) + \gamma_e^2B_{Gd}^2 S_{Gd}(\omega),
\end{equation}
where $\gamma_e$ is the NV's gyromagnetic ratio and $\omega =\omega_0$ in the absence of a static magnetic field.
Here $T_{1,bulk}$ describes the background NV relaxation, similar to that  in bulk diamond,  which does not depend on the ND size or other external factors. $B (S)$ is the noise strength (spectrum),  with the subscripts indicating   the contributions from  surface paramagnetic defects   or Gd molecules.
Our protocol sensitivity arises from 
the strong dependence of the $T_1$ time  on the surface density of Gd molecules $n$, as shown in Fig~\ref{fig:Fig2_singleNV}(a). 
A detailed derivation  of the $T_1$ model (see supporting material~\cite{SOM}) shows that this dependence is due to the change in the noise spectrum $S_{Gd}=R_{Gd}/(R_{Gd}^2+\omega_0^2)$  and strength, $B_{Gd}^2\sim n$.  Here $R_{Gd}\sim \sqrt n$ is the Gd fluctuation rate, mainly due to spin-spin couplings that are strongly distance-dependent. 
Thanks to  the strong   RNA-DNA hybridization compared to the DNA-PEI binding~\cite{SOM} we can assume that all the Gd molecules  get detached in the virus presence and  freely diffuse in the solution. Then, a higher initial c-DNA-DOTA-Gd$^{3+}$ density  yields a larger $T_1$ difference and a larger fractional change. We note that multiple Gd$^{3+}$ complexes can be incorporated into one c-DNA sequence~\cite{RotzACSnano2015} and the increased ratio between c-DNA and Gd$^{3+}$ complexes will lead to an even larger change in $T_1$.

These results demonstrate the potential for our proposed hybrid quantum sensor to not only detect the presence of the virus RNA, but even to 
quantitatively estimate the viral load with high sensitivity. This capability would 
allow  more accurately capturing the infectivity window and catching  more contagion cases -- even when the viral load is too small for other methods --  thus playing an important role in accurate estimation of epidemic trajectories. 
To quantify the sensor's performance, we thus consider the minimum number of detectable RNA copies in a total integration time $T$. This is related to the  \textit{quantum sensitivity}, $\delta n_{min}\sqrt{T}$, to variations in the surface c-DNA-DOTA-Gd$^{3+}$ density (see supporting material~\cite{SOM} for details). 
Figure~\ref{fig:Fig2_singleNV}(b)  shows the calculated  sensitivity for varying parameters and an  integration time $T=1$s. As expected, a smaller distance between NV and Gd$^{3+}$ complex (smaller ND diameter and Gd bound)  would result in a better sensitivity.
 
To  evaluate the practical sensor performance, we need to take into account variations in the NV sensor parameters that affect the sensitivity, such as the randomness in ND diameter $d$,  Gd$^{3+}$ surface density $n$, as well as position of the NV centers with respect to the origin~\cite{SOM}. 
These variations give rise to broad distributions of the sensitivity shown in Fig.~\ref{fig:Fig2_singleNV}(c), which predicts the probability that  a single hybrid sensor containing one NV center can detect a certain amount of RNA copies, without any pre-characterization. 
We remark that when the sensor sufficiently interacts with the virus in sample, the minimal detectable number of RNA copies in 1-second integration time can be as low as 100. This is well below the viral load for positive clinical samples, which typically has $10^5 - 10^6$ RNA copies per throat or nasal swab~\cite{WolferNature2020} without nucleic acid amplification. Therefore, in this single-NV scenario, our sensor would reach a ultralow FNR ($<$1\%) compared to the common RT-PCR diagnosis method which can have a FNR above 25\%~\cite{KucirkaAIM2020}.

\section*{Sensor performance with ensemble measurements}

While measuring a single ND before and after introducing virus RNA can yield a considerably large $T_1$ difference, as we have demonstrated above, this protocol suffers from low photon counts and the challenge of addressing the same single ND. Moreover, for single NDs, the detectable number of RNA copies is upper-bounded by the number of  c-DNA-DOTA-Gd$^{3+}$ molecules on the ND surface, which is in turn limited by the surface area and surface density of c-DNA-DOTA-Gd$^{3+}$. 
We thus analyze a more practical scenario where the fluorescence signal from an ensemble of NDs is measured after a fixed dark time $\tau$ following the initialization laser pulse, which allows inferring  the relaxation rate.
Each NV center in the ensemble is characterized by different parameters, including its random position inside the ND, and (normal) distributions of the ND diameters  and surface density of c-DNA-DOTA-Gd$^{3+}$ around their nominal values.  

\begin{figure*}
    \centering
    \includegraphics[width=0.9\textwidth]{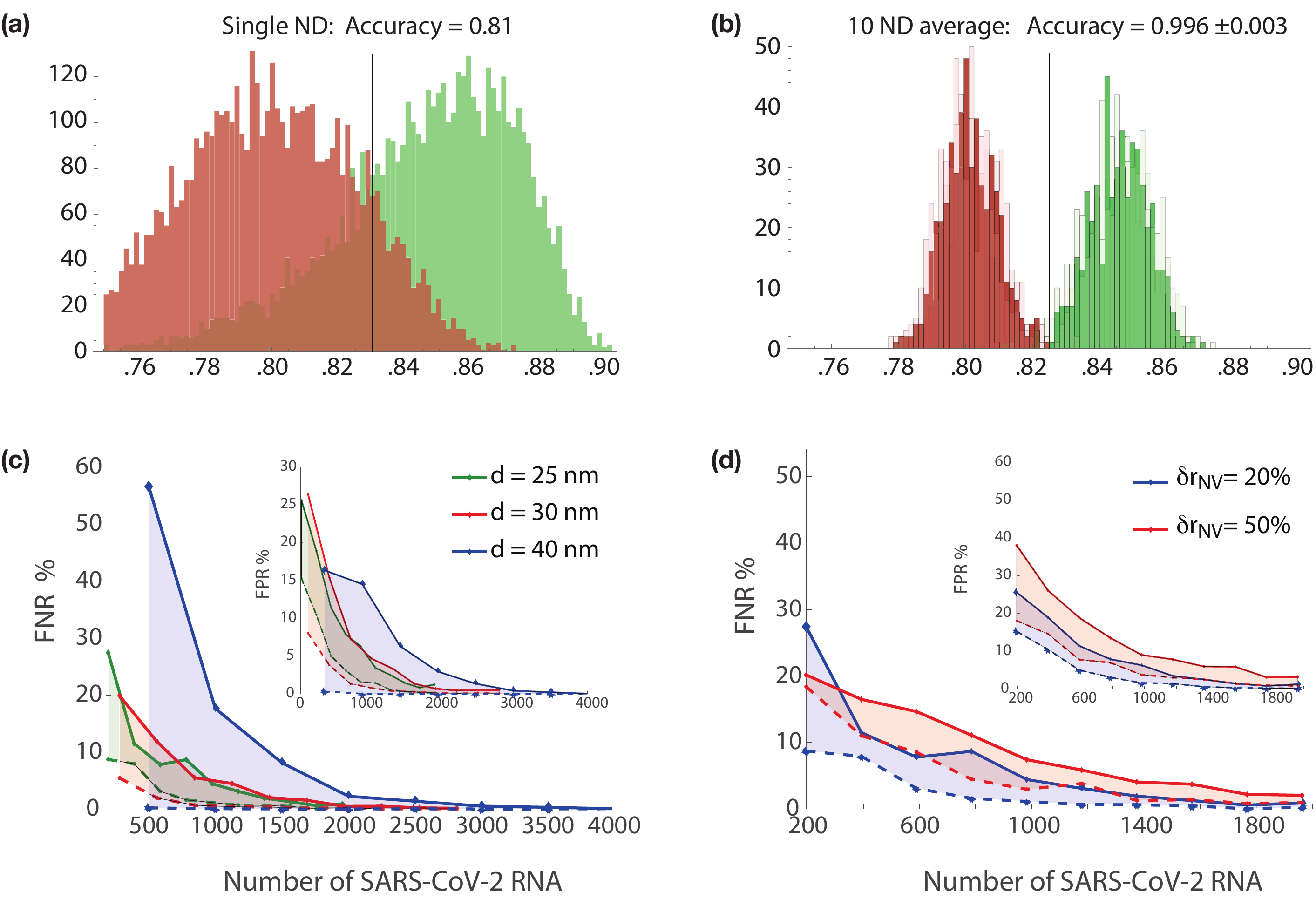}
    \caption{\textbf{NV ensemble sensor performance.} \textbf{(a)} Histogram of measured PL from single NDs at a fixed dark time $\tau$~=~200~$\mu s$ for $\bar{d}$ = 25~nm. The NV position is random in a sphere of 20\% of the ND radius. The red (green) distribution corresponds to the case where viral RNA is absent (present).  \textbf{(b)} The NDs in (a) are grouped into random ensembles of 10 NDs and averaged over. The histogram with deep (light) colors shows the PL without (with) photon-shot noise. In  both figures (a-b), the black lines indicate the optimal threshold that gives the maximum accuracy.
     \textbf{(c)} FNR (inset: FPR) as a function of number of SARS-CoV-2 RNA copies associated with ensembles of NDs with different diameters. The solid (dashed) curves show the worst (optimal) case where the photon-shot noise increases (decreases) the PL distribution overlap. The NV position is random in a sphere of 20\% of the ND radius. 
      \textbf{(d)} Same as (c), but here we compare the effect of reducing the uncertainty in the NV position from 50\% to 20\% of the ND radius (ND average diameter 25~nm). 
     In plotting the distributions in (c-d), we consider 5000 NDs with one NV each,  and 0.1~nm$^{-2}$ average surface c-DNA-DOTA-Gd$^{3+}$ density.  The ND diameter has a normal distribution with variance  3~nm and the average distance between ND surface and Gd molecules is 1.5~nm, with 0.2~nm variance.}
    \label{fig:Fig3_ensemble}
\end{figure*}

We then calculate the distribution of normalized photoluminescence (PL) counts at a  fixed waiting time $\tau$ for a large number of NDs (see Fig.~\ref{fig:Fig3_ensemble}). 
When the observed photon count is below a chosen threshold, indicating a large relaxation rate, we classify the result as  negative for the presence of the virus, while PL above the threshold indicate the virus presence. The FNR and FPRs arise when a misclassification occurs (e.g., the photon count is low even if virus RNA was present.) 
The threshold is found by maximizing the \textit{balanced accuracy} = 1- (FNR + FPR)/2 from the calculated PL  distributions, which describes the average of sensitivity (1-FNR) and specificity (1-FPR).  

In Fig.~\ref{fig:Fig3_ensemble}(a) we show how the PL arising from single NVs with/without the virus RNA is distributed. 
Due to the variation in ND diameters and other parameters, the two PL distributions arising from single NDs have a considerable overlap
, leading to FNR(FPR) = 0.14(0.239). To resolve this problem, a small number of NDs can be measured simultaneously,  leading to  well-resolved distributions. As few as 10 NDs can reach an accuracy $>$ 99.6 \% with FNR(FPR) $<$ 0.1(0.9)\%  (Fig.~\ref{fig:Fig3_ensemble}(b)). 
The potential FNR of our quantum sensor  is thus much lower than for the common RT-PCR method.  

To further show the performance of the quantum sensor in ensemble measurements, in Fig.~\ref{fig:Fig3_ensemble}(c) we present the FNR (FPR) as a function of the number of SARS-CoV-2 RNA copies that can be detected by a group of NDs. When a larger number of NDs are measured simultaneously, the PL distributions are more well-resolved, leading to lower FNR (FPR) as expected.  To take into account  the effects of photon-shot noise on  the PL distribution overlap we plot the two extreme cases: the photon shot noise can indeed either add or subtract to the (average) PL signal. Correspondingly,  the noise moves the signal PL distributions close to each other (worst case, shown in solid lines) or separate them further away (best case, shown in dashed lines). Even when considering photon shot noise, the FNR(FPR) achievable is still outstanding (only the largest diameter ND are considerably affected by the noise, since their average PL distributions are  narrower).

Further optimization of the ND parameters can lead to significant advantages, as already demonstrated in our simulations. For example, a large contributor to the dispersion in the PL curve of Fig.~\ref{fig:Fig3_ensemble}(a) is due to the random position of the NV in the ND, as in Fig.~\ref{fig:Fig3_ensemble}(d) where the FNR (FPR) for different distributions of NV position is shown. NV centers near the ND surface would also significantly suffer from random surface charge noise, potentially leading to deleterious charge dynamics of the NV centers.  This can be mitigated 
by better engineering  the NV-containing NDs (for example, with surface coating~\cite{AndreasABC2015,Neburkova2017}), to produce NV centers that are  well-below the ND surface and close to the NDs' origin. 
In addition, pre-characterization of the charge environment the NV centers would also help interpret the PL curves~\cite{BluvsteinPRL2019,GiriPRB2019}.


\section*{Discussions}
Until now we focused on variations in the quantum sensor properties that might limit the viral RNA detection. Indeed, we expect that  variations in other steps of the protocol, e.g., in the  detachment of Gd molecules due to the interaction between viral RNA and c-DNA on ND surface, will have a negligible influence. 
While other external factors could have been expected to induce the detachment even in the absence of viral RNA (increasing the  FPR), previous studies~\cite{ZhangACSnano2009,PetrakovaNanoscale2016,WuAFM2015} have demonstrated that the ND-PEI-DNA hybrid nanomaterial is a stable and efficient gene or drug delivery systems, and  survives both sonication and storage for several months. The binding between c-DNA-DOTA-Gd$^{3+}$ and ND surface should thus be stable against moderate temperature and mechanical fluctuations.
Furthermore, the hybridization of c-DNA and viral RNA is a highly efficient process~\cite{YangAC2008,LiuCC2009} and FNR induced by insufficient detachment of surface c-DNA-DOTA-Gd$^{3+}$ should be considerably small. 
%
Finally, to ensure sustained efficacy and specificity, it is imperative that the diagnosis methods target parts of the viral genome that are not considerably affected by naturally occurring viral mutations and are unique to SARS-CoV-2~\cite{Srinivasan2020}. Several methods that are based on sequence alignment currently exist to find the conserved parts of a viral genome (see for example~\cite{Rangan2020}). A generic text-mining method has been recently developed for rapid identification of segments of a whole genome that are likely to remain conserved during future genomic mutation events~\cite{Darooneh2020}.

Due to the rapid spreading of the pandemic, a high throughput diagnosis capacity is needed, which can be achieved by  our technique. Single or ensemble NDs can be incorporated in microfluidic devices with separated channels. Samples that possibly contain viral RNA are  injected into the channels and the resulting fluorescence signal is collected by a charge-coupled device camera. 
Although it's beyond the scope of this theory proposal, we note that practical issues such as bubbles or leakages in  microfluidic chips might slightly degrade the performance of the sensor. Alternative ways of mixing ND with the virus sample might be proposed to bypass the potential problems with microfluidic setup.
In addition to such simultaneous diagnosis of multiple samples, two other factors  ensure the scalability of our protocol:
Contrary to the general impression on the price of diamonds, synthesizing NDs that contain NV centers has become a mature technology and the material cost of single NDs is negligible. On the same note, NDs can make the technique more scalable than bulk diamonds with reduced cost. The system involves only short sequences of c-DNA and RNA, thus further limiting the cost in the chemical synthesis process.

The presented technology can be generalized to diagnosing other RNA virus such as HIV and MERS by using surface c-DNA that is specific to target virus. The technique can also be applied to detect DNA genome by replacing the c-DNA with an appropriate RNA sequence. While we consider NV centers in nanodiamonds, alternative quantum sensors or host materials might be adopted. For example, sensors based on silicon-vacancy centers in silicon carbide~\cite{NagyNC2019} might be developed.

\section*{Conclusion}
We propose a hybrid quantum sensor for  detecting the RNA of SARS-CoV-2 virus based on NV centers in nanodiamonds. We built a theoretical model  to describe the quenching of NV's relaxation time due to dipolar interactions between NV center and Gd$^{3+}$ molecules and to evaluate the sensor's performance. As the viral RNA detaches the Gd$^{3+}$, large changes in the NV photoluminescence yield a detection limit as low as several hundreds of viral RNA copies. The FNR can reach less than 1 \%, which is considerably lower than state-of-art RT-PCR diagnosis method. 
The present diagnosis method is scalable, fast and low-cost, which can meet the requirements of accurate estimation of epidemic trajectories and slowing down the current COVID pandemic.

\acknowledgements
This work was supported in part by the US Army Research Office through Grant W911NF-15-1-0548 and Canada First Research Excellence Fund.

\bibliographystyle{ieeetr}
\bibliography{manu_NL_revised}

\clearpage
\newpage


\section*{Supplementary material}

# SARS-CoV-2 quantum sensor based on nitrogen-vacancy centers in diamond

Changhao Li,[1,2] Rouhollah Soleyman,[3] Mohammad Kohandel,[3] and Paola Cappellaro[1,2,4]

[1]*Research Laboratory of Electronics, Massachusetts Institute of Technology, Cambridge, Massachusetts 02139, United States*
[2]*Department of Nuclear Science and Engineering,*
*Massachusetts Institute of Technology, Cambridge, Massachusetts 02139, United States*
[3]*Department of Applied Mathematics, University of Waterloo, Waterloo, Ontario, Canada*
[4]*Department of Physics, Massachusetts Institute of Technology, Cambridge, Massachusetts 02139, United States*
(Dated: November 9, 2021)

## CONTENTS

## I. CRISPR-BASED DIAGNOSIS

CRISPR technology could be integrated with our proposed quantum RNA sensor to achieve even better performance. CRISPR gene-editing technology has been used for virus RNA detection. These detection methods work by creating a guide RNA (gRNA) that is complementary to the target RNA sequence. Both Cas13- and Cas12-based systems have been developed for SARS-CoV-2 detection[1–5]. Binding of the guide RNA to complementary target RNA activates scissors-like function of the Cas-gRNA complex, cleaving any surrounding single-stranded RNA (ssRNA) –including any fluorescent RNA reporter, which confirms detection of the virus.

We anticipate that similar CRISPR-based tools can be used in conjunction with the ND-based quantum sensor. In this protocol, a ssRNA reporter connects $Gd^{3+}$-containing molecules to the ND. The activated Cas-gRNA complex cleaves the ssRNA reporter and detaches $Gd^{3+}$ from the ND surface, resulting in weaker magnetic interaction between $Gd^{3+}$ molecules and NV centers inside ND. In turns, this results in the fluorescent signal to increase above the virus-detection threshold. As explained in describing the previous protocol, the NV properties make the detection method based on magnetic noise particularly sensitive and robust.

CRSIPR-based detection can be made rapid and more sensitive, as shown in several recent studies. For instance, Fozouni et.al[6] recently developed a rapid Cas13-based assay that does not require pre-amplification of the viral genome. The authors showed that multiple gRNAs that target different regions of the viral RNA can be used to enhances sensitivity of the test. The method was able to detect about 100 copies per microliter in 30 minutes of measurement time and enabled accurate diagnosis of pre-extracted RNA from patient samples in 5 min. Therefore, the diagnosis method based on the integration between ND-based quantum sensing and CRISPR technique would be fast and sensitive.

## II. SENSITIVITY TO VARIATION OF SURFACE GD DENSITIES

In this section we derive the sensitivity of our protocol to changes in the surface density of Gd molecules, $n$, assuming a single NV center at the center of a nanodiamond. While some of the results below have been derived in reference[7,8], we briefly repeat them here for the convenience of readers.

We consider performing $N$ measurements on a single NV center for a total time $T$, each measurement detecting the spin relaxation after a time $\tau$ (we neglect dead times associated with polarization and detection of the NV centers, as these are small with respect to $\tau$, so that $T = N\tau$). The signal, that is, the total number of detected photons, is given by:

$$N(\Gamma_1) = \mathcal{D} T_D \frac{T}{\tau} \left(1 + C e^{-\Gamma_1 \tau}\right), \tag{1}$$

where $T_D = 500$ ns is the detection window, $\mathcal{D}$ the photon detection rate (we assume $\mathcal{D} = 10^5 s^{-1}$ here, which is typical for a single NV center) and $C \approx 0.2$ is the signal contrast for a typical single NV measurement. Here $\Gamma_1 = 1/T_1$ is the relaxation rate:

$$\Gamma_1 = \Gamma_1^{const} + 3\gamma_e^2 B_{\perp,Gd}^2 \frac{R_{Gd,tot}}{R_{Gd,tot}^2 + \omega_0^2}, \tag{2}$$

where $\Gamma_1^{const}$ is the (constant) relaxation rate contributions from bulk and surface spins[8].

The signal is photon shot-noise limited, that is, the noise is $N_{noise} \approx \sqrt{DT_D T/\tau}$.

### Variance of transverse magnetic noise

The transverse magnetic noise field can be calculated by summing up contributions from all Gd spins. It can be shown that[7,8]:

$$B_{\perp,Gd}^2 = \sum_i B_{\perp,i}^2 = \sum_i \left(\frac{\mu_0 \gamma_i \hbar}{4\pi}\right)^2 C_s \frac{2 + 3\sin^2 \theta_i}{r_i^6}, \tag{3}$$

where $C_{s,Gd} = \frac{S(S+1)}{3} = 5.25$ is determined by the Gd spin number $S$, $\gamma_i$ is the gyromagnetic ratio of spin $i$, $r_i$ is the distance between NV center and spin $i$ and $\theta_i$ is the angle between the NV quantization axis and the relative direction

of spin $i$ with respect to the NV spin. For a single NV center at the center of a ND, the strength of the magnetic noise is given by:

$$B_{\perp,Gd}^2 = \left(\frac{\mu_0 \gamma_e \hbar}{4\pi}\right)^2 C_{s,Gd} \int_0^{2\pi} d\phi \int_0^{\pi} (2+3\sin^2\theta)\sin\theta d\theta \lim_{\Delta r \to 0} \frac{\int_{d_0/2}^{d_0/2+\Delta r} \frac{r^2}{r^6} dr}{\Delta r n}$$
$$= \left(\frac{\mu_0 \gamma_e \hbar}{\pi}\right)^2 C_{s,Gd} \pi \frac{n}{(d_0/2+l)^4} \tag{4}$$

where $d_0$ is the diameter of ND, $l$ is the distance between ND surface and Gd complex and $n$ is the surface density of Gd molecules.

### Dipolar interactions between nearby Gd complexes

The magnetic noise induced by Gd spins is subject to fast fluctuations. Since the Gd spins are attached to the ND surface, the diffusion and rotation rates are negligible and the dipolar interaction $H_{ij}$ between nearby Gd spins $i$ and $j$ is the dominant term in the total fluctuation rate[8]. The dipolar fluctuation rate is given by[8]:

$$\hbar R_{Gd,dip} = \sqrt{\sum_{i \neq j} \langle H_{ij}^2 \rangle} = \frac{\mu_0 \gamma_e^2 \hbar^2 \sqrt{6} C_{s,Gd}}{4\pi} \left(\frac{\pi}{2}\right)^{1/2} \frac{n^{1/2}}{r_{min}^2} \Rightarrow R_{dip,Gd} \approx 21.1 \text{ns}^{-1} \text{nm} \times n^{1/2}. \tag{5}$$

where $r_{min}$ is the minimum allowed distance between surface spins. Here we take $r_{min} = 0.5$ nm, which is in the same order as the size of Gd complex[8]. Possible folding of c-DNA sequences and their interactions in three dimensional space might bring Gd spins close together[9]. We note that in our parameter regime where $n$ is on the order of 0.1 nm$^{-2}$, one can take $R_{Gd,tot} = R_{Gd,dip}$.

### Sensitivity calculation

A variation in the Gd surface density $\delta n$ will induce a change in the relaxation $\Gamma_1 \to \Gamma_1'$, thus changing the number of detected photons. If the variation is small, we can consider just the first order:

$$\Gamma_1' = \Gamma_1 + \delta\Gamma_1 = \Gamma_1 + \frac{3}{2}\gamma_e^2 \frac{B_{\perp,Gd}^2}{n} \frac{R_{Gd,dip}^3 + 3\omega^2 R_{Gd,dip}}{(R_{Gd,dip}^2 + \omega^2)^2} \delta n, \tag{6}$$

The associate signal variation is $\delta N = N(\Gamma_1) - N(\Gamma_1')$,

$$\delta N = \mathcal{D} T_D \frac{T}{\tau} C e^{-\Gamma_1 \tau}(1 - e^{-\delta\Gamma_1 \tau}) \approx \mathcal{D} T_D T C e^{-\Gamma_1 \tau} \delta\Gamma_1 \tag{7}$$

Since the photon shot noise dominates the experiments, we can take the noise to be $N_{noise} = \sqrt{N(\Gamma_1)} \approx \sqrt{\mathcal{D} T_D T/\tau}$, where we neglected contributions from $C$ as the contrast is typically small, $C \ll 1$. We thus obtain the signal-to-noise ratio (SNR):

$$\text{SNR} = \frac{\delta N}{\delta N_{noise}} = \sqrt{\mathcal{D} T_D T \tau} C \delta\Gamma_1 e^{-\Gamma_1 \tau} \tag{8}$$

The SNR is maximum at $\tau = T_1/2$,

$$\text{SNR}_{max} = \frac{\delta\Gamma_1}{\sqrt{\Gamma_1}} C \sqrt{\frac{\mathcal{D} T_D T}{2e}} \tag{9}$$

Then, the minimum fluctuation rate change $\delta R_{Gd}^{min}$ that can be detected is obtained for a SNR of one. We thus obtain the sensitivity per unit time:

$$\delta n_{min} \sqrt{T} = \frac{1}{C} \frac{\sqrt{2e\Gamma_1}}{\sqrt{\mathcal{D} T_D}} \frac{2n(R_{Gd,dip}^2 + \omega_0^2)^2}{3\gamma_e^2 B_{\perp,Gd}^2 (R_{Gd,dip}^3 + 3\omega_0^2 R_{Gd,dip})} \tag{10}$$

When the Gd molecules are the dominant relaxation source, we can neglect the constant contribution $\Gamma_1^{const}$ and simplify the sensitivity and reach Eq.1 in main text:

$$\delta n_{min}\sqrt{T} \approx \frac{n}{C\sqrt{\mathcal{D}T_D}}\sqrt{\frac{8eR_{Gd,dip}}{3\gamma_e^2 B_{\perp,Gd}^2}\frac{(R_{Gd,dip}^2+\omega_0^2)^{3/2}}{R_{Gd,dip}^3+3\omega_0^2 R_{Gd,dip}}}. \tag{11}$$

Then the minimal number of detectable cDNA-Gd pairs is given by

$$\delta N_{min}\sqrt{T} = (\delta n_{min}\sqrt{T})\pi d^2. \tag{12}$$

## III. CONTRIBUTION FROM FREE GD MOLECULES IN SOLUTION

Here we evaluate the noise contribution from the detached c-DNA-Gd molecules. Now the $T_1$ time of NV center is determined by

$$\frac{1}{T_1} \sim B_s^2 S_s(\omega) + B_{Gd,1}^2 S_{Gd,1}(\omega) + B_{Gd,2}^2 S_{Gd,2}(\omega), \tag{13}$$

where Gd,1(2) denotes the surface (free) Gd molecules. The goal is to evaluate the density of free Gd molecules.

After detachment from the nanodiamond surface, the c-DNA-DOTA-Gd$^{3+}$/RNA compound will diffuse freely in the solution. We consider the mean squared distance (MSD) of diffusion and calculate the lower bound of the volume of free diffusion

$$V = (6Dt)^{3/2} = (6\times 10^{-9}m^2s^{-1}\times 1s)^{3/2} \approx 4.6\times 10^{14}nm^3 \tag{14}$$

where we take the diffusion time $t=1$s and diffusion constant to be $D=10^{-9}m^2s^{-1}$.

We consider a 20 nm nanodiamond with surface density of c-DNA-Gd molecules being 1 $nm^{-2}$. After they are all detached from the surface, we estimate the free Gd density to be $n_2 \approx 10^{-5}nm^{-3}$. The contribution to the NV $T_1$ time from such a small Gd volume density is negligible.

## IV. BINDING REACTION IN THE SENSOR PROTOCOL

In the main text, we proposed to attach the c-DNA sequence to the ND surface via polyethyleneimine (PEI)[10,11], although other polymers or molecules such as poly-L-lysine (PLL)[12], polyvinylamine (PVA), and polyallylamine (PAA)[13], could be studied to improve efficiency. PEI can be easily adsorbed on oxygen-terminated NDs, since oxidized ND surfaces can form polar interactions such as hydrogen bonding and electrostatic interactions. PEI further binds to nucleic acids (NA) via non-covalent interactions[14], in particular electrostatic interactions due to opposing charge of PEI and NA. The resulting conjugates largely conserve the activity of the adsorbed biomolecules, thus allowing further functionalisation exploiting existing functional groups[15]. The ND-PEI-DNA complex has been shown to be efficient for gene or drug deliveries[14,15]. Molecular dynamic simulations[16] on the DNA-PEI interaction show that the binding time is several tens of nanoseconds with a typical binding constant $K_{eq}$ on the order of $10^5$ M$^{-1}$[17]. A separate study[18] estimates the optimal weight ratio between ND and nucleic acid strands, which we find corresponds to a number ratio of ND:c-DNA $\approx$ 1:100 for ND with d=20 nm. This is in agreement with the Gd density $n\approx 0.1$ nm$^{-2}$ that we use in our simulation. The attachment of c-DNA on ND surface should thus be efficient, ensuring the stability of the proposed sensor system.

To have a better understanding on the interaction between c-DNA and RNA, we perform thermodynamic simulation of their hybridization using the online UNAFold server[19,20]. As stated in the main text, to selectively detect the SARS-CoV-2 virus, we consider a nucleic acid sequence that is specific to the virus genome *5'-GGCU...CGG-3'* (labelled by sequence A) and its complementary DNA sequence *5'-CCGA...GCC-5'* (labelled by sequence B). As shown in Fig. 1, the melting temperature, which is defined as the point at which the concentration of double-stranded molecules $AB$ is one-half of its maximum value, is found to be $T_m \approx 80°C$ for the two selected sequences. This is well-above room temperature, at which the optical experiment is performed. Below $T_m$ the hybridized sequence $AB$ dominates the concentration with a mole fraction near one. Furthermore, we find that the difference between the ensemble free energies at low temperature (at $T_{min} < T_m$, where $AB$ is the main specie) and at high temperature (at $T_{max} > T_m$, where the duplex molecules disassociate and single-strand $A(B)$ sequences dominate the system) is $\Delta G \approx 53.6$ kcal/mol, with the corresponding ensemble entropy (enthalpy) difference being $\Delta S \approx 675.3$ cal/mol/K ($\Delta H \approx 237.3$ kcal/mol). It implies that under room temperature, the binding process occurs spontaneously with a high equilibrium constant.

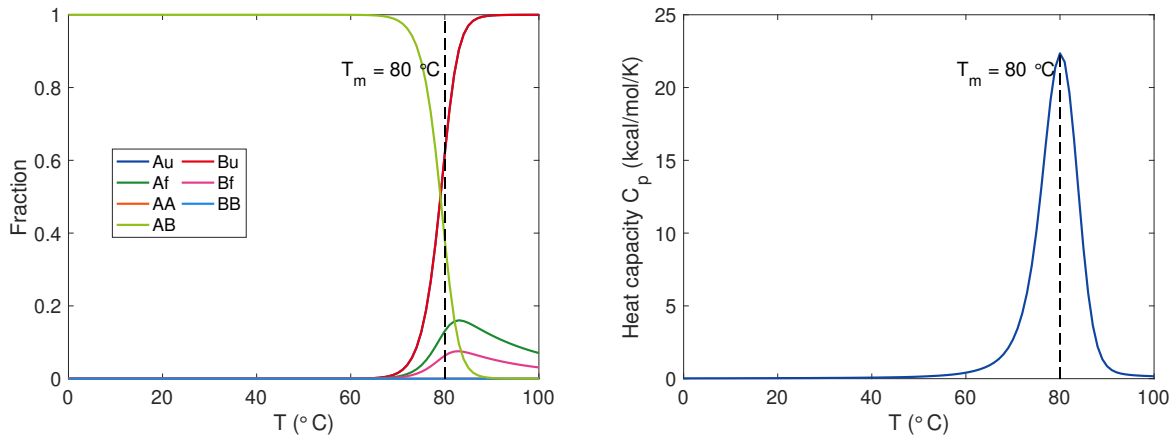

FIG. 1. Simulation of the binding between RNA (A) and c-DNA (B) sequence. *Left:* Plot of mole fractions of each species as a function of temperature T. Au (Bu) denotes unfolded sequence while Af (Bf) is for self-folding. The melting temperature is determined to be $T_m = 80°C$. *Right:* ensemble heat capacity $C_p$ as a function of temperature. A maximum point is shown near the melting temperature $T_m$. In the simulation the initial concentration for the sequences are $[A] = [B] = 10^{-5}$ M while the salt concentration is $[Na^+] = 1$ M.

As an example, at $40°C$ (a temperature higher than any rise in local temperature due to laser heating) the equilibrium constant for the reaction $A + B \Rightarrow AB$ is given by

$$K'_{eq} = \frac{[AB]}{[A][B]} \approx 6 \times 10^{19} M^{-1} \gg K_{eq} = \frac{[BC]}{[B][C]} \approx 10^5 M^{-1} \quad (15)$$

where $C$ ($BC$) denotes PEI (PEI-DNA). This implies that under equilibrium conditions, one would expect $\frac{[AB]}{[BC]} = \frac{K'_{eq}}{K_{eq}} \frac{[A]}{[C]} \gg 1$. This results would not substantially change even when considering other competing reactions (e.g. RNA attachment to PEI), given the many orders of magnitude difference.

To this end, we anticipate that the c-DNA/RNA binding will dominate over the interaction between c-DNA and PEI[16,17] on the nanodiamond surface. (Similarly, we do not expect the virus RNA to attach to the PEI.) Thus, we expect that the concentration of the double stranded molecule, detached from the PEI, will dominate over the residual c-DNA-PEI concentration.

---


\end{document}